\documentclass[prd,nofootinbib,preprint,superscriptaddress,twocolumn,10pt]{revtex4-2}
\pdfoutput=1
\usepackage[T1]{fontenc}
\usepackage{amsmath,amssymb}
\usepackage{braket}
\usepackage{epsfig}
\usepackage{graphicx}
\usepackage[usenames,dvipsnames]{xcolor}
\usepackage{subfigure}
\usepackage{slashed}
\usepackage[colorlinks,citecolor=blue]{hyperref}
\usepackage{comment}

\begin{document}


\title{High-Quality Axion Dark Matter at Gravitational Wave Interferometers}

\author{Disha Bandyopadhyay}
\email{b.disha@iitg.ac.in}
\affiliation{Department of Physics, Indian Institute of Technology Guwahati, Assam 781039, India}

\author{Debasish Borah}
\email{dborah@iitg.ac.in}
\affiliation{Department of Physics, Indian Institute of Technology Guwahati, Assam 781039, India}

\author{Nayan Das}
\email{nayan.das@iitg.ac.in}
\affiliation{Department of Physics, Indian Institute of Technology
Guwahati, Assam 781039, India}

\author{Rome Samanta}
\email{romesamanta@gmail.com}
\affiliation{Scuola Superiore Meridionale, Largo S. Marcellino 10, I-80138 Napoli, Italy}
\affiliation{  Istituto Nazionale di Fisica Nucleare (INFN), sez. di Napoli, Via Cinthia 9, I-80126 Napoli,
Italy}

\begin{abstract}

Gravitational effects are known to violate global symmetries, threatening the Peccei-Quinn (PQ) solution to the strong CP problem. Ultraviolet completions featuring a gauged $U(1)$ symmetry, where $U(1)_{\rm PQ}$ arises as an accidental global symmetry, can suppress Planck-suppressed operators, enabling high-quality axions in a mass window where it can also account for the observed dark matter (DM) in the Universe. We show that in such models, the spontaneous breaking of the $U(1)$ gauge symmetry generates a strong stochastic gravitational wave background (SGWB) from gauge cosmic string loops. Even in the most conservative scenario, for breaking scales $\gtrsim 10^{14}$ GeV, the SGWB signal strength can exceed astrophysical foregrounds across a broad frequency range.  Such quality axion models have a characteristic IR break frequency originating from the dynamics of the string-wall network collapse. We propose this characteristic SGWB frequency-amplitude region, identified as
\textit{Signature-Window-Axion-Gravitational waves} (SWAG), to be a novel probe of high-quality axion DM at future space and ground-based interferometers.
\end{abstract}
\maketitle
{ \bf  Introduction—} Stringent upper bounds on the neutron electric dipole moment (EDM) indicate that strong interactions are CP symmetric, even though the Standard Model (SM) as a whole is not. CP violation in the strong sector is parameterized by the angle $\bar{\theta}$, which is constrained from above as $\bar{\theta} < 10^{-10}$ \cite{Abel:2020pzs}. The requirement that $\bar{\theta}$ be so extremely small constitutes the \textit{strong CP problem}. A well-known solution is to promote $\bar{\theta}$ to a dynamical pseudoscalar field, the axion, emerging from the spontaneous breaking of a global $U(1)_{\rm PQ}$ symmetry, known as the Peccei-Quinn (PQ) mechanism~\cite{Peccei:1977hh, Peccei:1977ur, Wilczek:1977pj, Weinberg:1977ma, Preskill:1982cy, Kawasaki:2013ae,DiLuzio:2020wdo}. When the axion acquires a potential from non-perturbative QCD effects, $\bar{\theta}$ dynamically relaxes to zero, thereby solving the strong CP problem. In addition to this elegant resolution of the strong CP problem, axions can also solve other longstanding puzzles in particle physics like the origin of dark matter (DM) and baryon asymmetry of the Universe (BAU) \cite{Planck:2018vyg, ParticleDataGroup:2024cfk}. Axions can constitute all of cold dark matter if they are extremely light, with a mass of order $m_a \simeq 10^{-5}~\mathrm{eV}$ corresponding to axion decay constant $f_a\sim 10^{11}$ GeV, as suggested by cosmological considerations~\cite{Preskill:1982cy, Abbott:1982af, Dine:1982ah}. Depending on the details of the QCD axion model namely, Dine-Fischler-Srednicki-Zhitnitsky (DFSZ) \cite{Dine:1981rt, Zhitnitsky:1980tq} or Kim-Shifman-Vainshtein-Zakharov (KSVZ) \cite{Kim:1979if, Shifman:1979if}, additional contributions from the topological defects can alter this window by less than an order of magnitude \cite{Kawasaki:2013ae}. On the other hand, there are several ways in which axions can assist in generating the observed BAU \cite{Servant:2014bla, Ipek:2018lhm, Croon:2019ugf, Co:2019wyp}.

Despite its appeal, the PQ mechanism is challenged by explicit global symmetry-breaking operators~\cite{Kamionkowski:1992mf,Holman:1992us,Barr:1992qq,Ghigna:1992iv}, particularly those induced by quantum gravity~\cite{Kallosh:1995hi}, which is believed to violate global symmetries. Such effects can shift the axion potential minimum away from $\bar{\theta} \simeq 0$, destabilizing the strong CP solution. While one might expect Planck-suppressed operators like $\lambda \Phi^5 / M_{\mathrm{Pl}}$ to be harmless, dimensional analysis suggests otherwise. For $f_a \sim 10^{11}~\mathrm{GeV}$, astrophysically allowed~\cite{Raffelt:2006cw}, this operator competes with the QCD axion potential unless $\lambda \lesssim 10^{-40}$. The need for such extreme fine-tuning to preserve $\bar{\theta} < 10^{-10}$ defines the \textit{axion quality problem}.

To address the axion quality problem and construct high-quality axion theories, two broad strategies have been explored:

{\textit{Category-1: Suppressing gravity-induced operators}}--This class of models aims to eliminate or suppress dangerous Planck-suppressed operators. This can be achieved either by (a) generating a naturally small $\lambda$ via non-perturbative effects, e.g., $\lambda \sim e^{-S}$ with $S \gtrsim 200$~\cite{Abbott:1989jw,Coleman:1989zu,Kallosh:1995hi,Alvey:2020nyh}, or (b) enforcing gauge symmetries that forbid operators of the form $\Delta V \sim \lambda_n \Phi^n / M_{\mathrm{Pl}}^{n-4}$ up to large $n \gg 5$ with $\lambda_n \sim \mathcal{O}(1)$~\cite{Berezhiani:1989fp, Barr:1992qq,Babu:2002ic,Qiu:2023los,DiLuzio:2017tjx,Duerr:2017amf,Fukuda:2017ylt,Ibe:2018hir,Ardu:2020qmo,Babu:2024udi,Babu:2024qzb}. Composite axion models also provide natural suppression~\cite{Randall:1992ut,Lillard:2018fdt,Gaillard:2018xgk,Vecchi:2021shj,Lee:2018yak,Cox:2023dou,Cox:2021lii,Nakai:2021nyf,Contino:2021ayn,Podo:2022gyj}.

{\textit{Category-2: Strengthening infrared dynamics}}--An alternative approach introduces a second QCD-like sector with a higher confinement scale $\Lambda_{\mathrm{QCD}'}$, resolving the strong CP problem in both sectors simultaneously. These models typically predict heavy, cosmologically unstable axions~\cite{Holdom:1982ex,Treiman:1978ge,Flynn:1987rs,Gherghetta:2020keg,Berezhiani:2000gh,Dimopoulos:2016lvn,Hook:2019qoh}, and evade the usual astrophysical bounds, allowing $f_a \ll 10^8~\mathrm{GeV}$. Related ideas include multiple SM copies~\cite{Hook:2018jle,Banerjee:2022wzk}, enlarged QCD symmetries~\cite{Gherghetta:2016fhp}, and extra dimensions~\cite{Choi:2003wr,Reece:2025thc,Craig:2024dnl,Svrcek:2006yi}.

While experimental prospects for both categories remain broad and promising~\cite{Graham:2015ouw,FASER:2018eoc,Gligorov:2017nwh,Chou:2016lxi,Feng:2018pew,Chakraborty:2021wda,Bertholet:2021hjl,Kelly:2020dda}, there is growing interest in alternative detection channels such as the stochastic gravitational wave background (SGWB). Ref.~\cite{ZambujalFerreira:2021cte} studies SGWB signals from domain wall dynamics in \textit{Category-2} models, while first-order confinement transitions in the heavy QCD sector can lead to additional GW signatures~\cite{Dunsky:2023ucb}.

In this work, we demonstrate that even the simplest high-quality axion models within \textit{Category-1b} can naturally give rise to strong and detectable gravitational waves (GW). These GW are sourced by gauge cosmic strings (CS)~\cite{Kibble:1976sj,Hindmarsh:1994re}, which generically emerge from the ultraviolet (UV) completion of the theory featuring a $U(1)_X$ gauge symmetry \cite{Barr:1992qq,Babu:2024udi,Babu:2024qzb}. The CS network breaks in the early universe due to its interactions with domain walls (DW)~\cite{Barr:1992qq,Kawasaki:2013ae,Vilenkin:1982ks} formed around axion oscillation temperature. The overall GW spectrum exhibits a characteristic infrared (IR) frequency, here referred to as the Martin-Vilenkin (MV) frequency: $  f_{\rm IR}^{\rm break,\, MV}$ \cite{Martin:1996ea}.

We show that for a wide range of model parameters, the resulting GW background can exceed astrophysical foregrounds, making \textit{Category-1b} axion models a promising and testable framework for upcoming GW observatories such as LISA~\cite{2017arXiv170200786A}, DECIGO~\cite{Kawamura:2006up}, and Advanced LIGO (aLIGO)~\cite{LIGOScientific:2014pky}. In addition to their distinctive GW signatures, these models maintain axion stability compatible with dark matter, a feature not generally realized in earlier heavy axion + GW scenarios~\cite{ZambujalFerreira:2021cte,Dunsky:2023ucb}.

{\bf The framework—} Since our primary objective is to compute the GW spectra arising from gauged CS in generic \textit{Category-1b} axion models, the model-dependent details of the explicit particle content are not crucial. The key inputs are the $U(1)_X$ breaking scale and the DW formation scale at which the CS network dissolves. For concreteness, we consider a Babu–Dutta–Mohapatra (BDM) class of KSVZ-type axion models~\cite{Babu:2024qzb}, based on the gauge group $\mathcal{G}_{\rm SM} \times U(1)_X$ (see Appendix \ref{appen1} for more detail). In this setup, a minimum of four vector-like quarks ($Q_{iL,R}$, $i = 1\text{--}4$) and two singlet scalars ($T$, $S$) are required. Additionally, three chiral fermions ($N_{iR}$, $i = 1\text{--}3$) are introduced to ensure anomaly cancellation for the gauged $U(1)_X$. The global $U(1)_{\rm PQ}$ symmetry then emerges as an accidental symmetry of the Lagrangian. Alternatively, one could adopt the Barr–Seckel (BS) class of models~\cite{Barr:1992qq}, which also naturally realize the desired $U(1)_{\rm PQ}$ structure. 

 This setup naturally gives rise to two distinct symmetry-breaking scales: the gauge symmetry breaking scale $f_g \sim f_T$, and the global PQ symmetry-breaking scale $f_a \sim f_S$. Throughout our analysis, we assume a hierarchy $f_g \gg f_a$. The axion emerges as a linear combination of the phases of the scalar fields $S$ and $T$, effectively locking their phases to the axion potential. This phase locking is the key mechanism that causes both the gauge and global strings to become attached to domain walls once the axion potential is turned on~\cite{Barr:1992qq,Rothstein:1992rh,Babu:2024qzb,Babu:2024udi}. 

The lowest-dimensional operator that is gauge-invariant but explicitly violates the PQ symmetry, potentially induced by gravitational effects, takes the form ~\cite{Babu:2024qzb}:
\begin{equation}
   V_{\rm gravity} = \kappa\, e^{i \delta}\, \frac{S^{3n} T^\dagger}{(3n)!\, M_{\rm Pl}^{3n - 3}} + \text{h.c.}\label{vgrav}
\end{equation}

\begin{figure}
    \centering
    \includegraphics[width=0.85\linewidth]{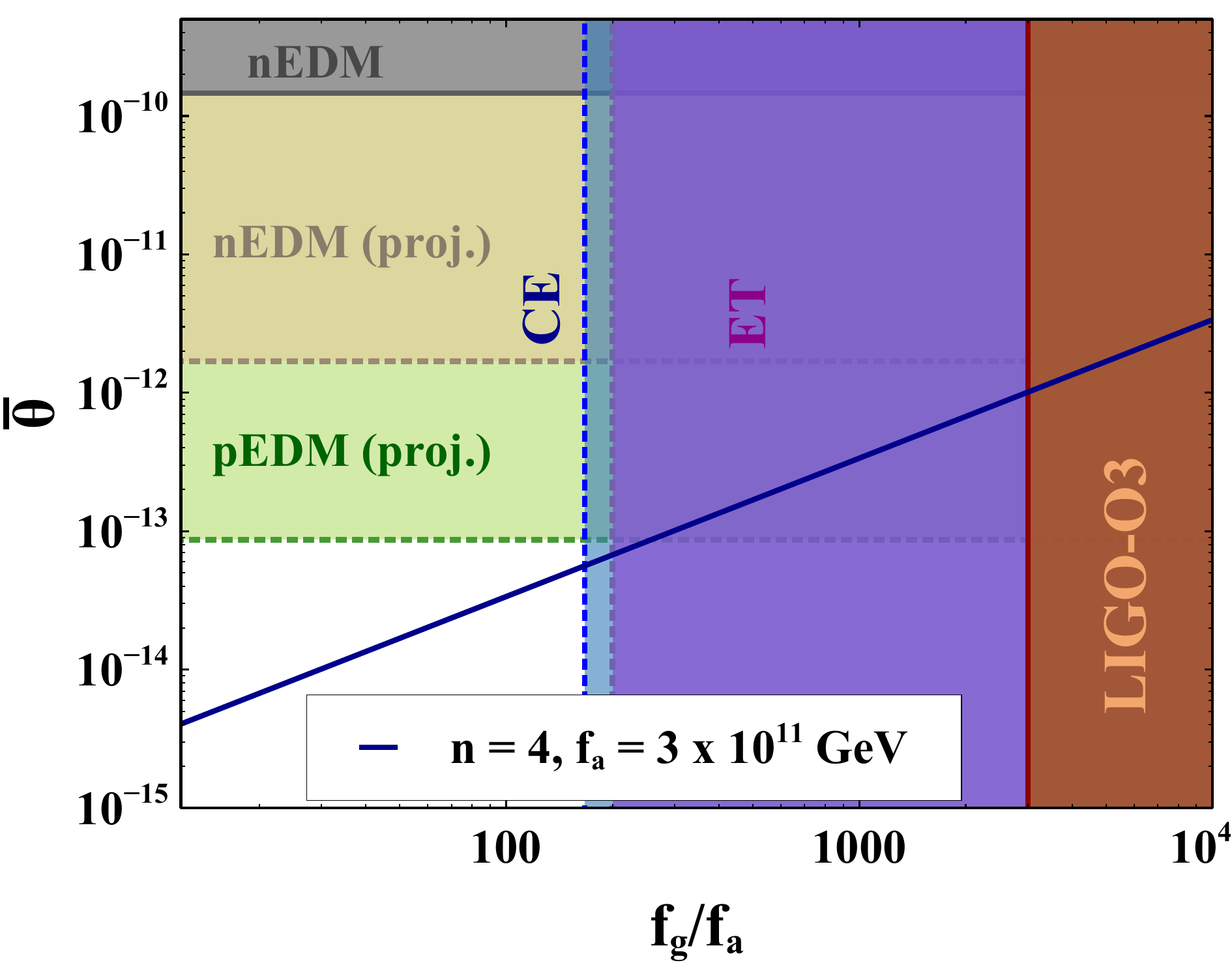} 
    \caption{$\bar{\theta}$ versus hierarchy of scales $f_g/f_a$, for $n = 4$ and $f_a = 3 \times 10^{11}$ GeV, which satisfies the relic abundance for axion being DM. The gray shaded region with a solid border is disfavored from current bound on neutron EDM (nEDM) \cite{Abel:2020pzs}, while the shaded regions with dotted borders denote projected future sensitivities for nEDM \cite{nEDM:2019qgk} and proton EDM (pEDM) \cite{Omarov:2020kws} respectively. The vertical shaded region with solid border is ruled out by LIGO-O3 \cite{KAGRA:2021kbb} observations while the ones with dotted borders are within reach of future GW experiments CE \cite{LIGOScientific:2016wof,Reitze:2019iox} and ET \cite{Punturo:2010zz} respectively.} \label{fig:fig1}
\end{figure}

Here $n$ is a positive integer related also to the $U(1)$ charges of the $S$ field, and the induced shift in the effective $\bar{\theta}$ parameter due to $V_{\rm gravity}$ can be estimated as (see Appendix \ref{appen1} for detail)
\begin{equation}
    \bar{\theta} \sim \frac{\kappa \,{\rm sin\delta} \,(3n)^{3n} f_a^{3n-2}f_g}{(3n)! \,2^{(3n-1)/2}\,M_{\rm Pl}^{3n-3} \, m_a^2}.
\end{equation}
where $m_a$ is the zero-temperature axion mass, fixed by the requirement that axions constitute the entirety of dark matter. Assuming the correct DM relic abundance arises from the vacuum misalignment mechanism~\cite{Preskill:1982cy,Abbott:1982af,Dine:1982ah}, one obtains $f_a \simeq 3 \times 10^{11}$~GeV and $m_a \simeq 17\, \mu$eV. \textcolor{black}{For simplicity, we take $\kappa \,{\rm sin\delta} = 1$ throughout our analysis.} It is then evident that the axion quality problem is resolved ($\bar{\theta} \ll 10^{-10}$) for large values of $n \geq 4$, as illustrated in Fig.~\ref{fig:fig1}.

Interestingly, for $n = 4$ and moderate values of $f_g / f_a$, the model predicts testable values of $\bar{\theta}$, potentially within reach of future neutron and proton EDM experiments. Furthermore, part of the parameter space is already constrained by LIGO-O3 upper limits on the stochastic GW background~\cite{KAGRA:2021kbb}, while the remaining parameter space can be probed by future GW interferometers such as CE~\cite{LIGOScientific:2016wof,Reitze:2019iox} and ET~\cite{Punturo:2010zz}, as discussed in the following sections.

In this class of models ~\cite{Babu:2024qzb,Barr:1992qq}, gauge CS form at a temperature scale $T \sim f_g$, while global CS form around $T \sim f_a$ \cite{Barr:1992qq,Rothstein:1992rh}. The gauge strings predominantly radiate gravitational waves~\cite{Matsunami:2019fss}, whereas global strings primarily lose energy through Goldstone boson emission~\cite{Baeza-Ballesteros:2023say}. A key feature of axion models is the formation of DW when the temperature-dependent axion mass becomes comparable to the Hubble parameter $H$, and the axion field begins to roll toward the minimum of its potential~\cite{Kawasaki:2013ae,DiLuzio:2020wdo}. The evolution and fate of the resulting string-wall network are governed by the domain wall number, $N_{\rm DW}$. For $N_{\rm DW} = 1$, the network is short-lived and efficiently dissipates its energy \cite{Vilenkin:1982ks}. In contrast, for $N_{\rm DW} > 1$, the network becomes stable and can eventually dominate the energy density of the Universe, leading to the well-known \textit{domain wall problem}. 
In high-quality axion models, this problem is  avoided if either class of strings, gauge or global, has a single domain wall attached. It has been shown in Refs.~\cite{Barr:1992qq,Rothstein:1992rh,Babu:2024udi,Babu:2024qzb,Ernst:2018bib} in detail that this condition can be satisfied in models with a $U(1)_{\rm local} \times U(1)_{\rm global}$ symmetry structure, such as the one considered here, making them cosmologically safe (see Refs.~\cite{Barr:1992qq,Rothstein:1992rh} for a comprehensive scholarly overview).

Even in the most conservative scenario, or from the GW detection standpoint, seemingly the worst-case scenario, where the gauge string network quickly dissolves through its interaction with DW, it survives long enough in the early Universe to emit a strong GW background from pure CS loops (neglecting, for now, contributions from other network remnants; we return to this later). Remarkably, this signal lies within the projected sensitivity range of upcoming interferometers. What sets high-quality axion models apart from generic gauge string–DW networks~\cite{Martin:1996ea} is that the DW formation temperature is fixed by the requirement of reproducing the observed DM relic abundance. Consequently, both the GW spectrum and the associated MV frequency are determined by a single parameter—the gauge string tension. This predictive feature is a central focus of this work.

{\bf Gravitational waves from gauge cosmic strings—} GW are radiated from cosmic string loops that are chopped off from long strings formed during the spontaneous breaking of the gauged $U(1)_{X}$ symmetry~\cite{Vilenkin:1981bx,Vachaspati:1984gt}. The long string network is characterized by a correlation length $L = \sqrt{\mu/\rho_\infty}$, where $\rho_\infty$ is the energy density in long strings and $\mu$ is the string tension. The tension is given by~\cite{Hill:1987qx}
$\mu = \pi f_g^2  h(\lambda', g')$, with $h(\lambda', g') \simeq 1$ for $\lambda' \simeq 2{g'}^2$, 
with $\lambda', g'$ being quartic self coupling of scalar field $T$ and $U(1)_X$ gauge coupling respectively. The evolution of a radiating loop with initial size $l_i = \alpha t_i$ is described by $l(t) = l_i - \Gamma G\mu (t - t_i)$, where $\Gamma \simeq 50$~\cite{Vilenkin:1981bx,Vachaspati:1984gt}, $\alpha \simeq 0.1$~\cite{Blanco-Pillado:2013qja,Blanco-Pillado:2017oxo}, $G = M_{\rm Pl}^{-2}$ is Newton’s constant, and $t_i$ is the time of loop formation.
\begin{figure*}
    \centering
    \includegraphics[width=0.45\linewidth]{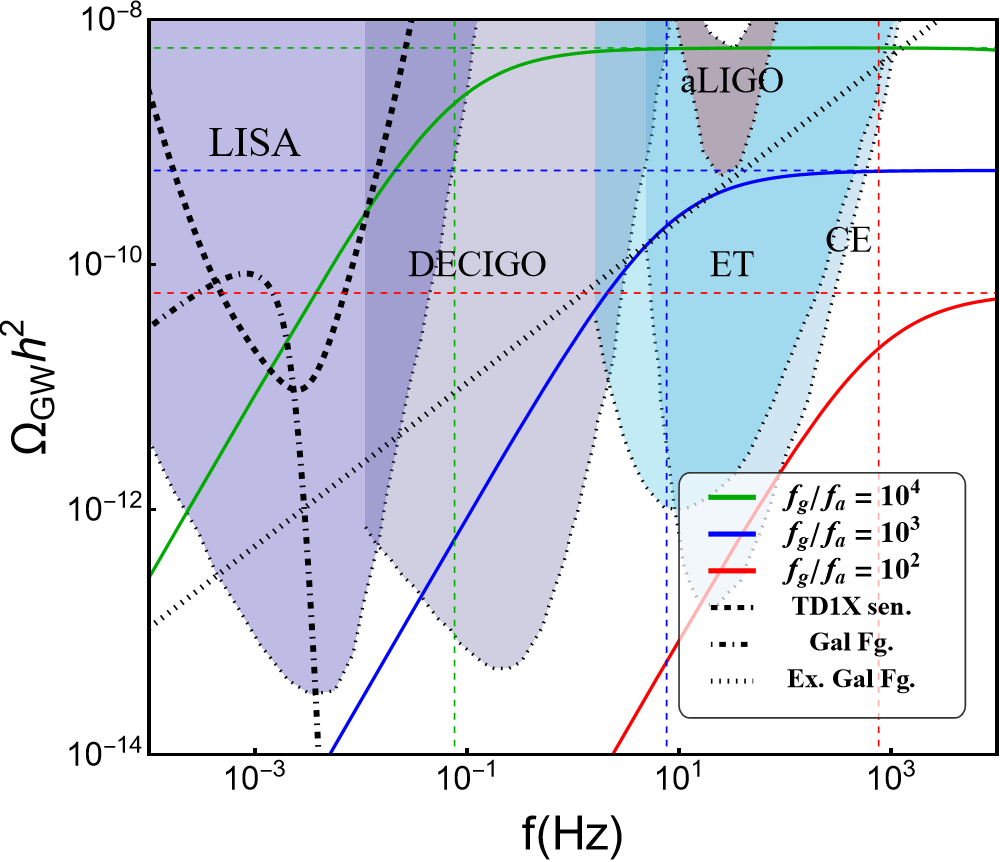}
    \includegraphics[width=0.455\linewidth]{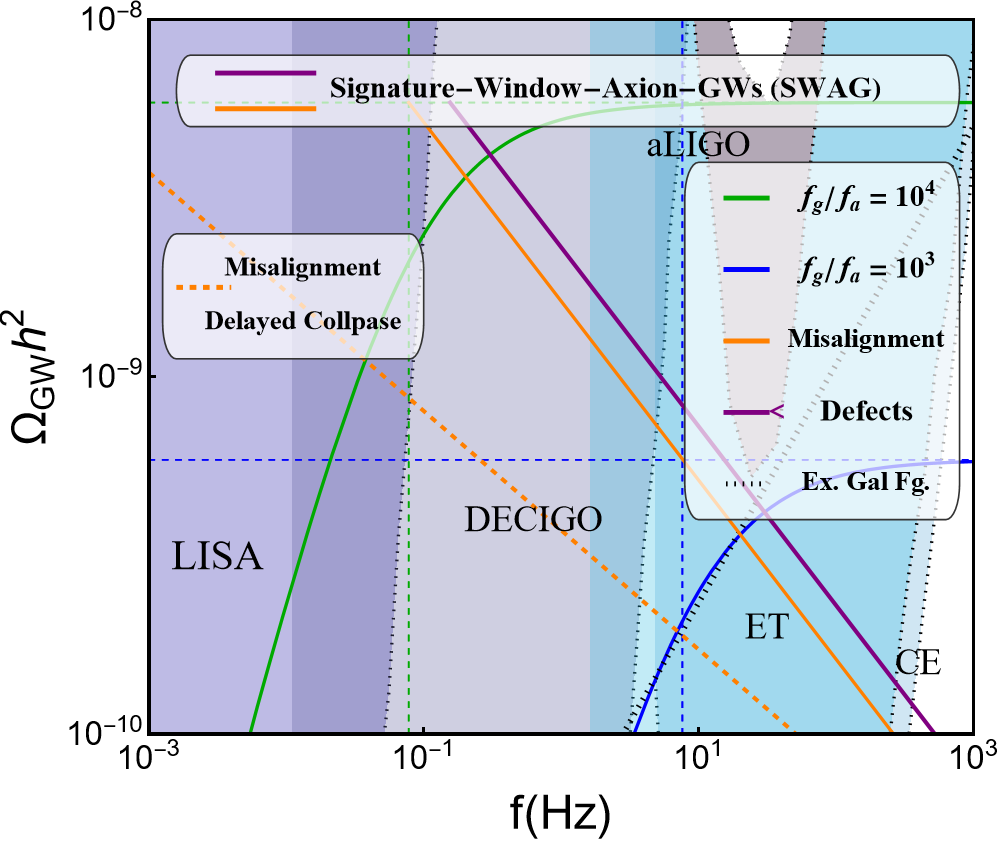}
    \caption{Left: Gravitational wave spectra (green, blue, red) for three benchmark values of $f_g / f_a$. The horizontal and vertical lines correspond to the relations defined in Eqs.~\eqref{flp1} and~\eqref{fbrk}. The dashed, dot-dashed, and dotted curves denote the LISA TD1X noise spectrum, galactic foreground, and extragalactic foregrounds, respectively. The white region overlapping with the aLIGO sensitivity represents the exclusion bound from LIGO-O3. Right: Zoomed-in view of the left panel. The region enclosed between the orange and purple lines defines the characteristic amplitude–frequency range associated with the SWAG, assuming the string network collapses at the domain wall formation temperature. The orange dashed line corresponds to a more realistic collapse time based on semi-analytic estimates. See main text for details.}\label{fig:figgw}
\end{figure*}
The total energy emitted in GW is distributed among normal-mode oscillations with frequencies
$ f_k = \frac{2k}{l_k} = \frac{a(t_0)}{a(t)} f $, where $k = 1, 2, \dots, k_{\rm max}$, $f$ is the observed GW frequency today, and $a(t)$ is the scale factor. The total GW energy density spectrum is given by~\cite{Blanco-Pillado:2013qja,Blanco-Pillado:2017oxo}
\begin{equation}
\Omega_{\rm GW}(f) = \sum_{k=1}^{k_{\rm max}} \frac{2k\, \mathcal{F}_\alpha\, G\mu^2\, \Gamma_k}{f\, \rho_c} \int_{t_i}^{t_0} \left( \frac{a(t)}{a(t_0)} \right)^5 n_\omega(t, l_k) \, \mathrm{d}t \,, \label{gwcs1}
\end{equation}
where $\rho_c$ is the critical energy density,   and $n_\omega(t, l_k)$ is the loop number density.

While the loop number density can, in principle, be computed using the velocity-dependent one-scale (VOS) model~\cite{Martins:1996jp,Martins:2000cs,Sousa:2013aaa,Auclair:2019wcv}, 
in this work, we adopt the loop number density from the numerical simulations of the Blanco-Pillado–Olum–Shlaer (BOS) \cite{Blanco-Pillado:2013qja}, which agrees with the VOS model for $\alpha,\mathcal{F}_\alpha \simeq 0.1$. As such, the gravitational wave spectra presented in our figures are strongly supported by numerical simulations, and so is the determination of the MV frequency. The power emitted in the $k$-th harmonic mode is given by
$\Gamma_k = \frac{\Gamma, k^{-\delta}}{\zeta(\delta)}$, 
where $\delta = 4/3$ for loops with cusps~\cite{Damour:2001bk}, and $\zeta(\delta)$ denotes the Riemann zeta function.

In general, the GW spectrum arising from a stable string network exhibits two prominent features:  
\textit{(i)} a low-frequency peak arising from GW radiation emitted by loops that were formed during the radiation-dominated epoch but decayed in the subsequent matter-dominated (standard) era, and  
\textit{(ii)} an approximately scale-invariant plateau at higher frequencies~\cite{Blanco-Pillado:2013qja,Blanco-Pillado:2017oxo,Sousa:2020sxs}, given by
\begin{equation}
\Omega_{\rm GW}^{\rm plt} = \frac{128\pi \mathcal{F}_\alpha G\mu}{9\, \zeta(\delta)} \, \frac{A_R}{\epsilon_R} \, \Omega_R \left[ (1 + \epsilon_R)^{3/2} - 1 \right] \,, \label{flp1}
\end{equation}
which originates solely from loop dynamics during the radiation-dominated era. In Eq.~\eqref{flp1}, $\epsilon_R = \alpha / \Gamma G\mu \gg 1$, $A_R \sim 5.4$ for radiation domination, and $\Omega_R \simeq 9 \times 10^{-5}$ is the present-day radiation energy density fraction. Importantly, the amplitude of the plateau scales as $\Omega_{\rm GW}^{\rm plt} \propto \sqrt{\mu} \propto f_g$, indicating that higher symmetry-breaking scales result in a stronger GW signal.

For quality axion models, the feature \textit{(i)} is absent, since the wall-string network dissipates well before matter-radiation equality (redshift $z \sim 3400$). Consequently, the spectrum exhibits only the feature \textit{(ii)}, with a blue tilt $\Omega_{\rm GW} \sim f^{3/2}$ for $f \lesssim f_{\rm IR}^{\rm break,\, MV}$ and a plateau for $f \gtrsim f_{\rm IR}^{\rm break,\, MV}$ (see, e.g., Fig. \ref{fig:figgw}). Using Eq.\eqref{gwcs1}, the condition for axion oscillation $3 \,H(T_{\rm osc}) \approx m(T_{\rm osc})$, and the power-law parametrisation of temperature dependent axion mass $m(T)$ \cite{DiLuzio:2020wdo,mt1,mt2}, we obtain an approximate analytical expression for the $f_{\rm IR}^{\rm break,\, MV}$ as
\begin{equation}
   f_{\rm IR}^{\rm break,\, MV}=0.7\left(\frac{10^{15}\, \rm GeV}{f_g}\right)^2\left(\frac{m_a}{30\, \mu \rm eV}\right)^{1/6}\, {\rm Hz}.\label{fbrk}
\end{equation}
Note that while Eqs.~\eqref{flp1} and \eqref{fbrk} are the two key expressions describing the GW spectrum in high-quality light axion models, the characteristic frequency $f_{\rm IR}^{\rm break, MV}$ arises from the near-instantaneous cessation of GW emission from pure string loops at the axion oscillation temperature $T_{\rm osc}$.  Although a precise treatment of the transition to late-time string dominance and a full computation of emission from wall–string systems would require lattice field theory simulations, semi-analytic estimates, such as those presented in Ref.~\cite{Dunsky:2021tih} (see also other relevant works related to string-DW collapse scenarios: \cite{Gelmini:2023ngs,Gouttenoire:2023gbn,Ghosh:2025cxp,Maji:2023fba,Lazarides:2023ksx,Maji:2025yms,Fu:2024rsm,Fu:2025dlp} ), can significantly improve the GW detection prospects by lowering $f_{\rm IR}^{\rm break, MV}$, as we will discuss shortly.  

 \textbf{Results and discussions—} In Fig.~\ref{fig:figgw} (left), we show three representative GW spectra corresponding to $f_g / f_a = 10^{4,3,2}$ (green, blue, red), with $f_a \simeq 3 \times 10^{11}$~GeV and $k = 1$. The signal amplitude cannot be arbitrarily large, as it is bounded from above by the non-observation of a stochastic GW background in LIGO-O3. \textcolor{black}{Similar upper bounds exist from pulsar timing array (PTA) observations and successful big bang nucleosynthesis (BBN) although they are not relevant for the range of frequency and amplitude shown in Fig.~\ref{fig:figgw}}. By summing over a large number of modes and imposing the LIGO-O3 bound at 25 Hz~\cite{KAGRA:2021kbb}, we find that $f_g \lesssim 9 \times 10^{14}$~GeV. Using Eq.~\eqref{fbrk}, this implies an absolute lower bound on the turning frequency: $f_{\rm IR}^{\rm break,\, MV} \gtrsim 0.7$~Hz, assuming that axions constitute $100\%$ of dark matter generated from vacuum misalignment. For smaller values of $f_g$, the overall signal amplitude decreases, while $f_{\rm IR}^{\rm break,\, MV}$ increases accordingly.

 The dashed, dot-dashed, and dotted curves in Fig.~\ref{fig:figgw} represent the LISA TD1X noise spectrum, the galactic foreground, and the extragalactic foregrounds, respectively. The galactic foreground arises primarily from unresolved white dwarf binaries within the Milky Way~\cite{Nissanke:2012eh,Evans:1987qa,Cornish:2017vip,Karnesis:2021tsh}, while the extragalactic foreground receives contributions from neutron stars, stellar-origin black hole binaries during their inspiral phase, and white dwarf binaries \cite{Regimbau:2011rp,Babak:2023lro,Lehoucq:2023zlt}. Notably, signals corresponding to $f_g \gtrsim 10^{14}$~GeV (typically, the region within the blue and the green curve in Fig.~\ref{fig:figgw}) can potentially avoid contamination from these astrophysical foregrounds, making them cleaner and higher-quality probes of primordial origin. Furthermore, even though LISA may barely detect the spectral turnover in such cases, the power-law part of the signal is sufficiently strong (the green curve) to rise above the instrumental noise.

As discussed in Refs.~\cite{Kawasaki:2014sqa,Benabou:2024msj,Gorghetto:2018myk}, when axion radiation from global strings and domain walls are included, the axion dark matter mass can be as large as $500~\mu$eV. In this case, the corresponding turning frequency increases beyond $f_{\rm IR}^{\rm break,\, MV} \gtrsim 0.7$~Hz (cf. Eq.~\eqref{fbrk}). This motivates the definition of a highly predictive \textit{Signature-Window-Axion-Gravitational waves} (SWAG), which characterizes the GW signal in high-quality light axion models. As shown in Fig.~\ref{fig:figgw} (right), the SWAG region lies between the orange ($m_a\sim 17~\mu$eV) and purple ($m_a\sim 500~\mu$eV) lines, and any $(\Omega_{\rm GW},\, f_{\rm IR}^{\rm break,\, MV})$ point within this region exhibits the characteristic spectral behavior: a rising spectrum $\Omega_{\rm GW} \propto f^{3/2}$ for $f \lesssim f_{\rm IR}^{\rm break,\, MV}$, followed by a flat plateau $\Omega_{\rm GW} \simeq \Omega_{\rm GW}^{\rm plt}$ for $f \gtrsim f_{\rm IR}^{\rm break,\, MV}$ \cite{Buchmuller:2021mbb,Sousa:2020sxs}. We propose SWAG as the most conservative prediction for high-quality light axion models. In its simplest form, SWAG is particularly robust, as it is simulation consistent and the spectral turnover is directly linked to the correct dark matter relic density sourced by axions. While one may argue that similar turnover features could emerge in other BSM scenarios, reproducing a SWAG-like window would typically require fine-tuning of model parameters. Notably, in most such cases, the domain wall formation temperature remains a free parameter, unlike in high-quality axion models, where it is fixed.

That said, several effects could influence the robustness of SWAG. First, as noted earlier, pure string domination may persist beyond the DW formation time. As discussed in Ref.~\cite{Martin:1996ea}, the string domination time can be characterized by $t_c \simeq \mu / \sigma$, where $\sigma$ is the axion DW tension. If $T_c < T_{\rm osc}$, the pure string network effect survives longer, resulting in a downward shift of the frequency $f_{\rm IR}^{\rm break,, MV}$. Following the method of Ref.~\cite{Dunsky:2021tih}, we find that the modified turnover frequency takes the form:
\begin{equation}
\overline{f}_{\rm IR}^{\rm break,\, MV} \simeq f_{\rm LT} \left( \frac{10^{15}~\mathrm{GeV}}{f_g} \right) f_{\rm IR}^{\rm break,\, MV},
\end{equation}
where the dimensionless late-time correction factor is $f_{\rm LT} \sim 10^{-2}$. This downward shift in frequency is illustrated by the dashed orange line in Fig.~\ref{fig:figgw} (right). Furthermore, if one includes contributions from remnant networks, such as walls bounded by strings formd around $t_c$, we find that $f_{\rm LT} \sim 10^{-6}$, a remarkable result that could shift the turnover frequency into the PTA sensitivity range. However, such an optimistic outcome requires confirmation from numerical simulations of the wall–string network.

In any case, we believe that our work motivates several additional promising directions for future investigation. These include, e.g., a detailed assessment of the robustness of the minimal SWAG prediction in the presence of astrophysical foreground contamination~\cite{Caprini:2019pxz,Caprini:2024hue,Samanta:2025jec,Blanco-Pillado:2024aca,Giese:2021dnw,Kume:2024xvh}. Moreover, it is possible to render the phase of the field $T$ blind to the axion potential via an appropriate charge assignment. In such a case, gauge strings are expected to radiate GW up to the present epoch, enhancing the GW signal at lower frequencies. While this scenario generally shifts the IR turnover frequency to lower values—resulting in a GW spectrum that seemingly lacks  the low-energy characteristic features of high-quality axion models at interferometer scales, it can open up the possibility of generating a UV spectral break. This behavior resembles the features discussed in Refs.~\cite{Cui:2018rwi,Gouttenoire:2019kij,Borah:2022byb,Chianese:2024gee}, and can arise if one of the heavy quarks dominates the energy density of the Universe~\cite{Cheek:2024ofn,Cheek:2023fht,Cheek:2025gvx}.

{\bf Conclusion—} That gauge strings can efficiently radiate GW is now a well-recognized phenomenon—acknowledged not only in theoretical studies but also in recent PTA data analyses \cite{EPTA:2023fyk,Reardon:2023gzh,Xu:2023wog,NANOGrav:2023hvm}, and embedded within the future physics roadmaps of experiments like LISA \cite{Auclair:2019wcv} and LIGO \cite{LIGOScientific:2021nrg}. Yet, it is somewhat surprising that this powerful feature has remained largely unexplored in the context of high-quality light axion models with gauged $U(1)$ symmetries, which are now gaining extensive attention. In this work, we demonstrate that such models can naturally predict a distinctive and detectable stochastic GW background characterized by an infrared spectral break (referred to
as the Martin-Vilenkin frequency) at the interferometer scales—a sharp contrast to the ultraviolet breaks typically studied in literature~\cite{Vilenkin:1991zk,Cui:2018rwi,Gouttenoire:2019kij,Borah:2022byb,Datta:2020bht,Blasi:2020wpy,Chianese:2024gee,Antusch:2024ypp}. Remarkably, the entire signal is governed by a single parameter: the $U(1)$ breaking scale (neglecting other theoretical uncertainties pertinent to signal reconstruction \cite{Blanco-Pillado:2024aca}). While previous studies on GW are focused primarily on high-quality heavy axions~\cite{ZambujalFerreira:2021cte,Dunsky:2023ucb}, we show that high-quality light axions not only generate similarly strong GW signals, but also remain fully consistent with the observed DM abundance. This elevates light axions to a compelling, multi-messenger target, testable both via interferometric GW detectors and axion search experiments (details given in Appendix \ref{appen2}), offering a powerful new window into the light, high-quality axion DM models.

{\bf Acknowledgments—} We thank K. S. Babu for insightful discussions related to high-quality axion models. We thank Rinku Maji for useful comments. The work of D.B. is supported by the Science and Engineering Research Board (SERB), Government of India grant MTR/2022/000575 and the Fulbright-Nehru Academic and Professional Excellence Award 2024-25. The work of N.D. is supported by the Ministry of Education, Government of India via the Prime Minister's Research Fellowship (PMRF) December 2021 scheme. R.S. acknowledges the support of the project TAsP (Theoretical Astroparticle Physics) funded by the Istituto Nazionale di Fisica Nucleare (INFN).

\appendix
{\color{black}
\section{More details on the model}
\label{appen1}
\subsection{Field content and charge assignment}
In the manuscript, we followed the notation of~\cite{Babu:2024qzb} and labeled the operators by the parameter $n$, which may appear ambiguous at first glance. However, $n$ should be understood as encoding the corresponding gauge charges. The notation introduced in~\cite{Barr:1992qq}, namely $(p,q)$, may indeed make this point more transparent. The central idea is to construct gauge-invariant operators characterized by the parameter $n$ (or equivalently $(p,q)$), and then appropriately choose the associated charges in order to avoid the axion quality problem.

\begin{table}[h!]
    \centering
    \begin{tabular}{|c|c|c|c|c|}
     \hline \hline Fields  & $U(1)_p$ &  $U(1)_q$ & $U(1)_X$ & $U(1)_{\rm PQ} $ \\
      & & & $\equiv U(1)_{p+q}$ & $\equiv U(1)_{p-q}$\\
     \hline   $Q_{iL}$ & $0$ & $1$ & 1 & -1\\
       $Q_{iR}$ & $0$ & $-1$ & -1 & 1\\
       $Q_{4L}$ & -3 & $0$ & -3 & -3\\
       $Q_{4R}$ & $3$ & $0$ & 3 & 3\\
       $N_{1R}$ & $2$ & $0$ & 2 & 2\\
       $N_{2R}$ & $4$ & $0$ & 4 & 4\\
       $N_{3R}$ & $-6$ & $0$ & -6 & -6\\
       $S$ & $0$ & $\frac{2}{n}$ & $\frac{2}{n}$ & $-\frac{2}{n}$\\
       $T$ & $6$ & 0 & 6 & 6\\
       \hline
    \end{tabular}
    \caption{New field content and quantum numbers under $U(1)_p \times U(1)_q$ and $U(1)_X \times U(1)_{\rm PQ}$ symmetries.}
    \label{tab1}
\end{table}
For the given charge assignment, it is evident that the lowest-dimensional operator that is gauge-invariant but explicitly violates the PQ symmetry, takes the form:
\begin{equation}
   V_{\rm gravity} = \kappa\, e^{i \delta}\, \frac{S^{3n} T^\dagger}{(3n)!\, M_{\rm Pl}^{3n - 3}} + \text{h.c.}\label{vgrav}
\end{equation}
In principle, values of $n<4$ are not forbidden by the gauged symmetry; however, for the axion decay constant required to obtain the observed relic abundance from vacuum misalignment, $f_a \simeq 3\times 10^{11}\,\text{GeV}$, such choices do not adequately solve the axion quality problem. As shown in Fig.~\ref{fig:n}, only for $n \ge 4$ the induced shift $\bar\theta$ remain well below $10^{-10}$.\\

\begin{figure}
    \centering
    \includegraphics[width=0.8\linewidth]{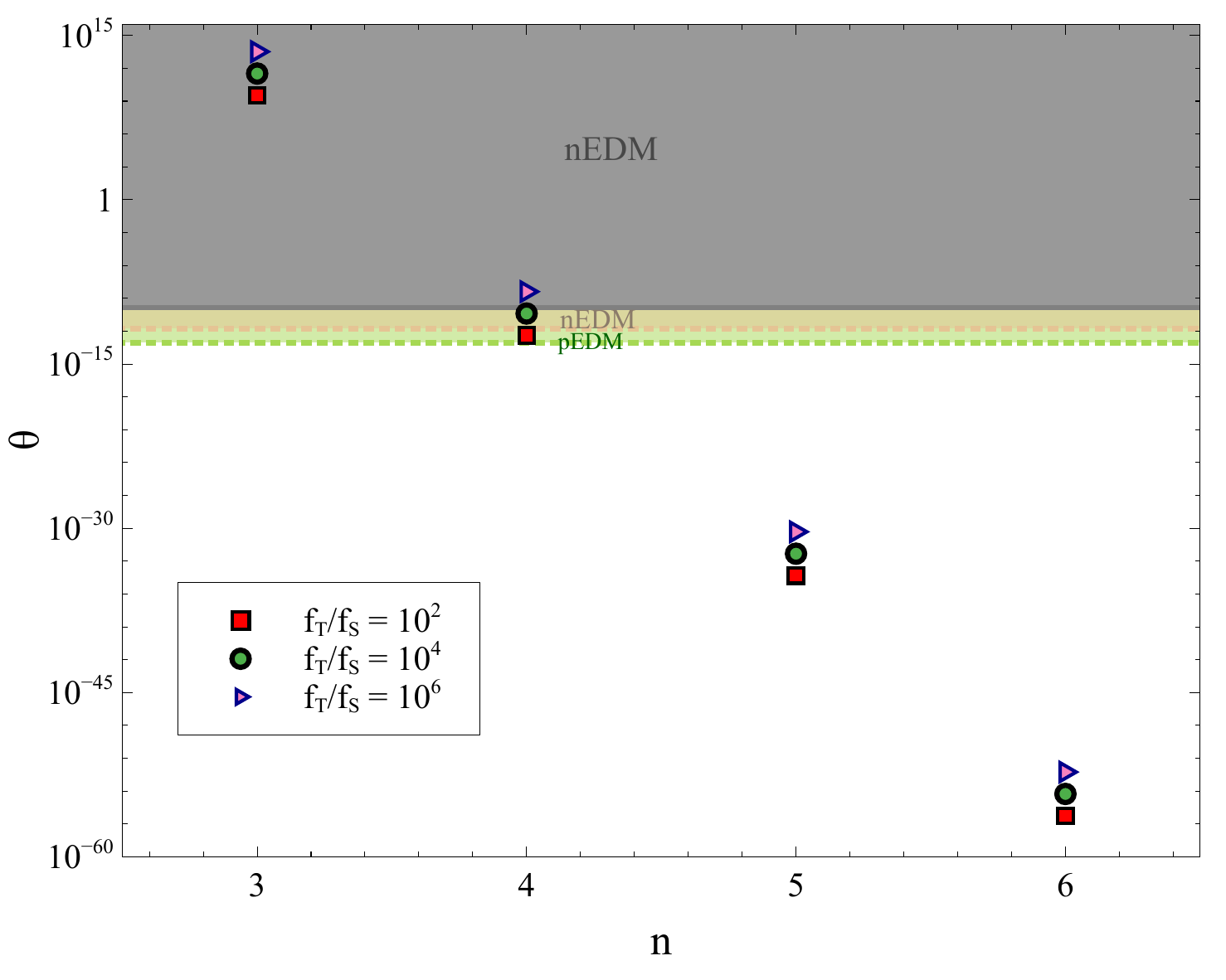}
    \caption{$n$ Vs. $\theta$, keeping $f_a\sim f_S\sim 3\times 10^{11}\,\text{GeV}$.}
    \label{fig:n}
\end{figure}

\subsection{Additional relics}

None of the newly introduced particles in Table~\ref{tab1} lead to unwanted cosmological relics. 
The vector-like quarks $Q_i$ and $Q_4$ can mix with the SM quarks. 
Similarly, the neutral fermions $N_{iR}$ can mix with the SM leptons, ensuring their eventual decay. 
For example, in addition to the leading-order terms responsible for generating the masses of these fermions (see the BDM model~\cite{Babu:2024qzb}), we can have higher-dimensional operators such as
\begin{align}
-\mathcal{L} \supset\; &
\overline{Q_{iL}} d_R \frac{S^2}{M_{\rm Pl}}
+ \overline{Q_{4L}} Q_{iL} \frac{S^4}{M^3_{\rm Pl}}
\nonumber \\[4pt]
&+ \overline{Q_{4L}} Q_{iL} N_{2R} N_{3R} \frac{1}{M^2_{\rm Pl}}
+ \overline{L}\tilde{H}N_{3R} \frac{T}{M_{\rm Pl}}
\nonumber \\[4pt]
&+ N_{3R} N_{2R} \frac{S^{*4}}{M^3_{\rm Pl}} ,
\label{eq1}
\end{align}
together with
\begin{align}
-\mathcal{L}_Y \supset\; &
Y_{44} \overline{Q_{4L}} Q_{4R} T^\dagger
+ \sum_i Y_{ij} \overline{Q_{iL}} Q_{jR}
\frac{S^n}{n!\Lambda^{\,n-1}}
\nonumber \\[4pt]
&+ y_{12} \overline{N^c_{1R}} N_{2R} T^\dagger
+ \frac{y_{33}}{2} \overline{N^c_{3R}} N_{3R}
\frac{T^2}{2!\Lambda}
+ {\rm h.c.}
\end{align}

In Eq.~\eqref{eq1}, we have assumed $n=4$ and taken the hypercharge of $Q_i$ to be the same as that of the SM down-type quarks. 
For the symmetry-breaking scales considered in this work, $f_g > f_a = 3 \times 10^{11}\,\mathrm{GeV}$, the operators in Eq.~\eqref{eq1} ensure that the newly introduced fermions decay into SM states well before the onset of BBN. 
Similarly, the new physical scalars can decay via Higgs-portal or Yukawa interactions into SM degrees of freedom. 

Therefore, requiring the lifetimes of the new particles to satisfy $\tau \ll 1\,\mathrm{s}$ does not impose any additional constraints on the model. 
\subsection{Domain-wall number and GW signals}

In this class of models PQ symmetry is not uniquely defined, since any linear combination of the accidental PQ symmetry and the gauged $U(1)_a$ also constitutes a valid PQ symmetry. 
The axion field can be written as
\begin{equation}
a = \sum_i c_i \, \eta_i 
  = \frac{3 n\, f_T \eta_S - f_S \eta_T}
  {\sqrt{9 n^2 f_T^2 + f_S^2}} \, ,
\end{equation}
from which the domain-wall number is determined as~\cite{Babu:2024qzb}
\begin{equation}
N_{\rm DW} 
= \min_{\, n_i \in \mathbb{Z}} 
\left( \frac{1}{f_a} \sum_i n_i \frac{c_i}{f_i} \right) .
\end{equation}
Substituting the explicit expressions for $c_S$, $c_T$, and $f_a$ relevant for our setup, this reduces to
\begin{equation}
N_{\rm DW} = \min \left( 3 n_1 - n_2 \right) .
\end{equation}
By choosing the integers $n_1 = 1$ and $n_2 = 3n - 1$, one can always construct a realization with $N_{\rm DW} = 1$. Although values $N_{\rm DW} > 1$ are more generic in this class of models, let us show that this regime is at best only marginally viable, since removing the long-lived string-wall network typically requires an explicit PQ-breaking term large enough to endanger axion quality.

For \(N_{\rm DW}>1\), the string-wall network can be long-lived, and an explicit PQ-breaking term is needed to lift the vacuum degeneracy and trigger annihilation. Omitting the overall \(\kappa\) coefficient and the \(\mathcal O(1)\) phase-dependent factor, the typical size of the gravity-induced bias is
\begin{equation}
\epsilon_{\rm bias}
\sim
2\,
\frac{f_S^{3n}f_T}{(3n)!\,2^{(3n-1)/2}M_{\rm Pl}^{3n-3}}\,
\sin\!\left(\frac{\pi}{N_{\rm DW}}\right).
\label{eq:eps_bias_Xi}
\end{equation}

Assuming that the network quickly enters the scaling regime, the annihilation epoch can be estimated from
\begin{equation}
H_{\rm ann}\sim \frac{\epsilon_{\rm bias}}{\sigma},
\end{equation}
where \(\sigma\) is the axion domain-wall tension. Assuming radiation domination, the corresponding annihilation temperature is
\begin{widetext}
\begin{equation}
 \frac{T_{\rm ann}^{\rm (RD)}}{\rm MeV}
=
1.3\,
\left(\frac{g_*}{10.75}\right)^{-1/4}
\left(\frac{|\bar{\theta}|}{10^{-10}}\right)^{1/2}
\left(\frac{f_a}{10^{11}\,{\rm GeV}}\right)^{-1/2}\\
\left[2\sin\!\left(\frac{\pi}{N_{\rm DW}}\right)\right]^{1/2},
\end{equation}
\end{widetext}
where we used
\begin{equation}
\bar{\theta}\simeq \frac{f_S^{3n}f_T}{(3n)!\,2^{(3n-1)/2}M_{\rm Pl}^{3n-3}m_a^2 f_a^2}.
\end{equation}
If the bias is too small, the wall network may dominate the energy density before annihilating. Estimating the onset of domination from \(\rho_{\rm DW}\sim \rho_R\), again assuming radiation domination, gives
\begin{equation}
\frac{T_{\rm dom}}{\rm MeV}
\simeq
0.8\,
\left(\frac{g_*}{10.75}\right)^{-1/4}
\left(\frac{f_a}{10^{11}\,{\rm GeV}}\right)^{1/2}
\left(\frac{6}{N_{\rm DW}}\right).
\label{eq:Tdom_num}
\end{equation}

Requiring the network to annihilate before it dominates, \(T_{\rm ann}^{\rm (RD)} > T_{\rm dom}\), implies the lower bound
\begin{equation}
|\bar{\theta}|
\gtrsim
3.4\times 10^{-10}\,
\left(\frac{f_a}{3\times 10^{11}\,\mathrm{GeV}}\right)^2
\left(\frac{6}{N_{\rm DW}}\right)^2
\frac{1}{2\sin\!\left(\pi/N_{\rm DW}\right)}.
\label{eq:thetaeff_lower_general_3e11}
\end{equation}
For \(N_{\rm DW}=\mathcal{O}(1\!-\!10)\), this lower bound is typically close to, or above, the experimental upper limit on \(\bar{\theta}\). Therefore, for \(N_{\rm DW}>1\) there is only a narrow window in which the explicit PQ-breaking term is large enough to eliminate the wall network early, yet small enough to preserve axion quality.

Few more comments are in order. In this work we primarily consider local strings and therefore anticipate that the infrared peak frequency may shift to lower values depending on the gauge symmetry-breaking scale. 
Local strings can efficiently radiate gravitational waves even for relatively small symmetry-breaking scales. 
By contrast, for global strings with breaking scale below $10^{15}\,\mathrm{GeV}$, the dominant emission channel is into Goldstone bosons rather than gravitational waves. 
In our setup, a comparatively smaller global string scale is in fact preferred in order to reproduce the observed axion abundance.

Within this framework, it is possible to engineer network decay by realizing $N_{\rm DW}=1$ for either type of string \cite{Rothstein:1992rh}. 
For example, one may obtain $N_{\rm DW}=1$ for global strings while having $N_{\rm DW}>1$ for local strings. 
Alternatively, an appropriate charge assignment (different from what is shown in Table \ref{tab1}) can render the local strings effectively insensitive to the axion potential, such that network annihilation must be seeded by global strings. 
In that situation, a characteristic gravitational-wave background can arise, for instance, through a modified expansion history induced by heavy quarks. On the other hand, in regions of parameter space with a standard cosmological expansion history, the resulting gravitational-wave spectrum reduces to the conventional local cosmic-string spectrum, without a distinctive imprint from the axion-sector structure. 
While phenomenologically viable, such a spectrum would be less readily distinguishable.}
\subsection{Axion quality} 
 Using chiral perturbation theory, the zero-temperature QCD axion potential can be derived as 
\begin{figure*}
    \centering
    \includegraphics[width=0.4\linewidth]{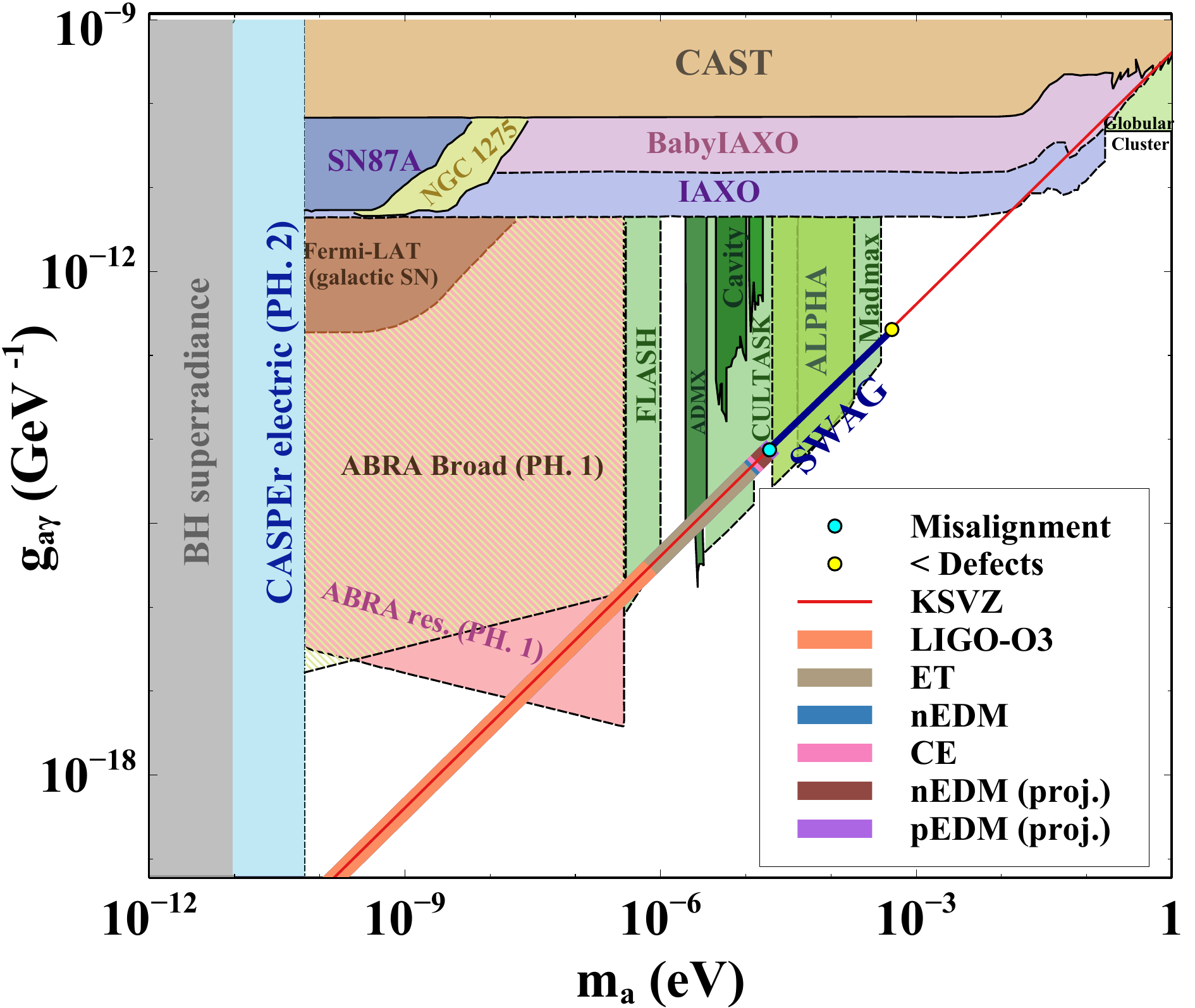}
    \includegraphics[width=0.4\linewidth]{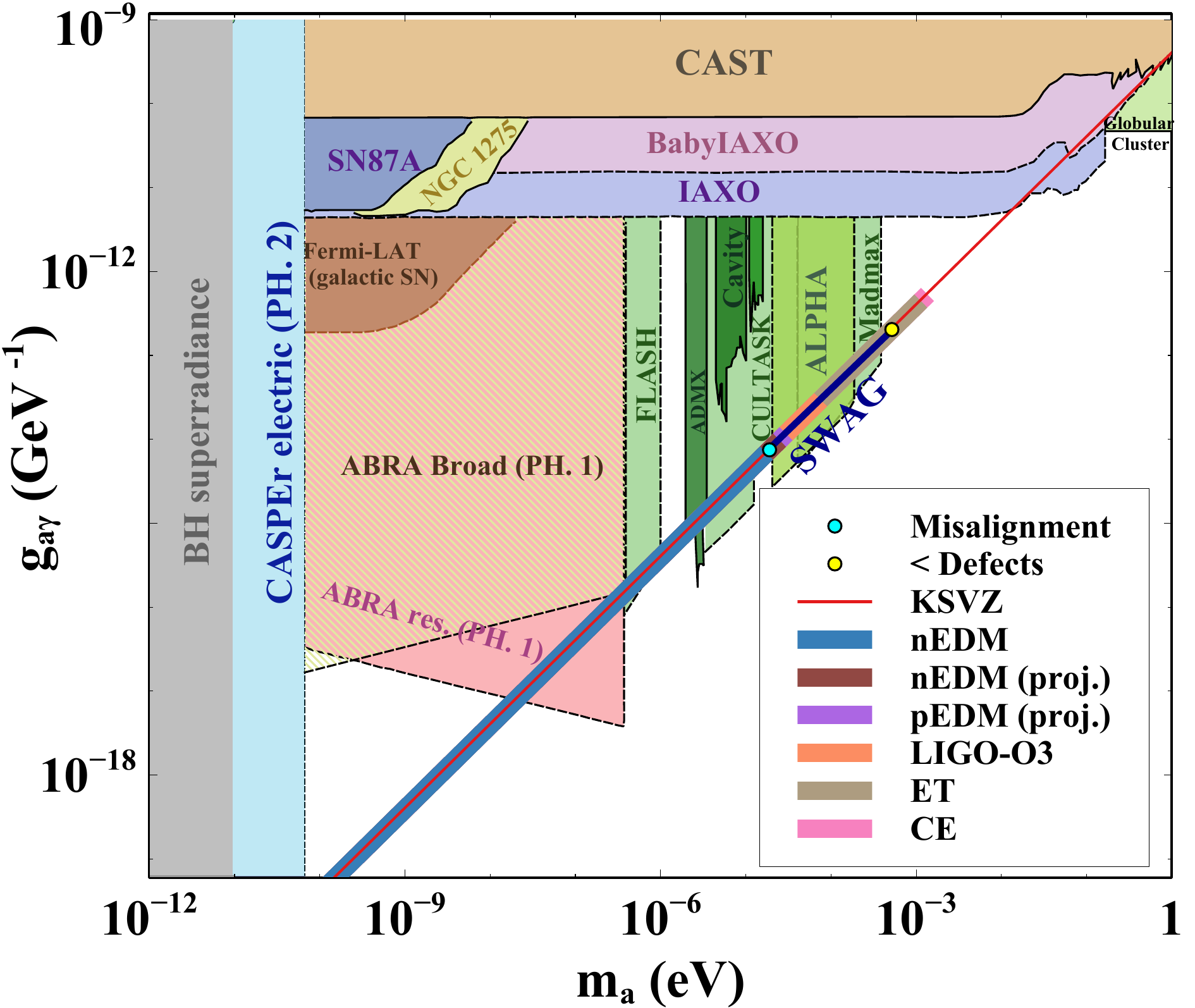}
    \caption{Axion-photon coupling $g_{a\gamma}$ versus axion mass $m_a$ for  $n=4$ and $f_g/f_a =10^2$ (left) and $f_g/f_a =10^4$  (right). The Cyan colored point denotes the point where the axions, generated by vacuum misalignment, accounts for entire dark matter abundance. The yellow colored point correspond to the maxium value of axion mass consistent with observed relic of DM, if produced from topological defects. The window between these two points is identified as SWAG corresponding to the unique GW signature of our model. The colored bands along the KSVZ model-predicted $g_{a\gamma}$ correspond to current bounds on neutron nEDM \cite{Abel:2020pzs}, SGWB (LIGO-O3) \cite{KAGRA:2021kbb} and future sensitivities for nEDM \cite{nEDM:2019qgk}, pEDM \cite{Omarov:2020kws}, SGWB (CE~\cite{LIGOScientific:2016wof,Reitze:2019iox} and ET~\cite{Punturo:2010zz}) respectively. Other solid and dashed contours represent different bounds and future sensitivities respectively for axions, see text for details.}
    \label{fig:app1}
\end{figure*}
\begin{equation}
       V \, (a) \, = \,-  f_\pi^2 m_{\pi}^2 \, \sqrt{1 \, - \, \dfrac{4 m_d m_u}{(m_u \, + \, m_d )^2} \, \sin^2 \left(\dfrac{a}{2 f_a} + \dfrac{\bar{\theta}}{2}\right) }.
\end{equation}
At high temperatures, we will get temperature-dependent mass and potential for axion. The $\bar{\theta}$ in the potential can be absorbed by transforming the axion field as $a \rightarrow a + f_a \bar{\theta}$. Using this potential, the axion mass below QCD scale can be found as \cite{Weinberg:1977ma}

\begin{equation*}
     m_a^2 = \dfrac{m_d m_u}{(m_u \, + \, m_d )^2} \dfrac{m_\pi^2 f_\pi^2}{f_a^2} ,   
\end{equation*}
\begin{equation}\label{axion mass}
\implies\,  m_a \, \approx \, 5.7 \, \left(\dfrac{10^{12} \, {\rm GeV}}{f_a}\right)\, {\rm \mu  eV}.
\end{equation}
The potential $V(a)$ is at minimum when
\begin{equation}
    \sin \left(\dfrac{a}{f_a} + \bar{\theta}\right) = 0.
\end{equation}
This minima is shifted when the gravity-induced PQ breaking higher dimensional operators are introduced which only respects the gauge symmetry. The lowest- dimensional PQ breaking but gauge invariant potential involving the scalars $S, T$ of our model is of the form:
\begin{equation}
   V_{\rm gravity}= \kappa e^{i \delta}\, \frac{S^{3n}T^\dagger}{(3n)! \,M_{\rm Pl}^{3n-3}}+\text{h.c.}
\end{equation}
Using the parametrisation of the scalar fields   $S = \frac{1}{\sqrt{2}}f_S\, e^{i \eta_S/f_S}, T =\frac{1}{\sqrt{2}} f_T\, e^{i \eta_T/f_T}$, the corresponding new contribution to the axion potential can be written as, 
\begin{equation}
   V_{\rm gravity}^{(a)}= \kappa\frac{f_S^{3n}f_T}{(3n)! \,2^{(3n-1)/2}\,M_{\rm Pl}^{3n-3}} \, \cos \left(\frac{a}{f_a}+\delta\right).
\end{equation}
Here, $ a = \frac{3 n\, f_T \eta_S - f_S \eta_T}{\sqrt{9 n^2 f_T ^2 + f_S^2}}$ and $f_a = \frac{f_S f_T}{\sqrt{9 n^2 f_T ^2 + f_S^2}}$. After taking into account the gravity-induced correction to the total axion potential which is now a sum of $V_{\rm gravity}^{(a)} + V (a)$ and minimizing it, the total shifted value of the QCD $\bar{\theta}$-vacuum is given as
\begin{equation}
    \bar{\theta} \sim \frac{\kappa \,{\rm sin\delta} \, f_S^{3n}f_T}{(3n)! \,2^{(3n-1)/2}\,M_{\rm Pl}^{3n-3} \, m_a^2 f_a^2}.
\end{equation}
In the limit $f_T>> f_S$, the axion decay constant is given by $f_a = \frac{f_S}{3n}$. This automatically ensures the hierarchy $f_g \gg f_a$ (with $f_g = f_T$), provided that $f_T\geq f_S/3n$. In this regime, the total shift of the QCD $\bar{\theta}$
-vacuum takes the form
\begin{equation}
    \bar{\theta} \sim \frac{\kappa \,{\rm sin\delta}\,(3n)^{3n} f_a^{3n-2}f_g}{(3n)! \,2^{(3n-1)/2}\,M_{\rm Pl}^{3n-3} \, m_a^2}.
\end{equation}
For suitable choices of symmetry breaking scales and the parameter $n$, this shifted value can be kept within experimental upper limits from neutron EDM \cite{Abel:2020pzs}.

\section{Axion detection experiments}
\label{appen2}
The axion detection prospects in most experiments primarily rely on axion-photon coupling. The axion-photon interaction is parametrized as 
\begin{equation}
    \mathcal{L}_{a\gamma} \supset \frac{1}{4} g_{a \gamma } F_{\mu \nu} \tilde{F}^{\mu \nu},
\end{equation}
where, $g_{a \gamma } = (\frac{E}{N} - 1.92)\frac{\alpha}{2 \pi f_a}$ \cite{Srednicki:1985xd} with $E, N$ representing electromagnetic and color anomaly coefficients, respectively. In our case, we choose vanishing $U(1)_Y$ charge of heavy quarks $Q_{L,R}$ which gives $ g_{a \gamma }$ to be same as the KSVZ model. Fig. \ref{fig:app1} shows the parameter space in the plane of axion-photon coupling $g_{a\gamma}$ and axion mass $m_a$ including the bounds and sensitivities of different experiments or observables. The line labeled as KSVZ refers to the model's predicted axion-photon coupling. The current experimental bounds on the axion-photon coupling from various experiments or observables are shown by the solid color lines (CAST \cite{CAST:2017uph}, SN87A \cite{Turner:1987by, Burrows:1988ah, Raffelt:1987yt}, NGC 1275 \cite{Fermi-LAT:2016nkz}, ADMX \cite{ADMX:2006kgb, ADMX:2019uok}, Globular clusters \cite{Ayala:2014pea}) whereas future experimental sensitivities or observables are shown by the dashed lines (CASPEr \cite{Budker:2013hfa}, FLASH \cite{Alesini:2023qed}, ABRACADABRA \cite{Ouellet:2018beu}, CULTASK \cite{Lee:2020cfj, Semertzidis:2019gkj}, MADMAX \cite{Caldwell:2016dcw}, IAXO \cite{IAXO:2019mpb},  Fermi-LAT \cite{Meyer:2016wrm}, ALPHA  \cite{PhysRevLett.123.141802, PhysRevD.107.055013}, black hole (BH) superradiance \cite{Cardoso:2018tly}).

The cyan colored point is consistent with the observed relic abundance of dark matter assuming the production via vacuum misalignment mechanism. If the PQ symmetry is to be broken in post-inflationary era, then the initial value of the misaligned axion field $\theta_i = \pi/\sqrt{3} \simeq 1.81$, which results in the observed DM abundance $\Omega_{a}{\rm h}^2=0.12$ for $f_a \, \sim \, 3.3 \times 10^{11} \,$ GeV and axion mass $m_a \sim 17 \, \mu$eV. The yellow colored point correspond to the maximum value of axion mass up to which its relic can be generated from topological defects like global CS and DW \cite{Kawasaki:2014sqa,Benabou:2024msj,Gorghetto:2018myk}. Similar to the right panel of Fig. \ref{fig:figgw}, here we identify the range $m_a \in 17-500 \, \mu$eV as the SWAG window with the unique GW signal of our high-quality axion DM model. The other colored bands along the KSVZ-predicted $g_{a\gamma}-m_a$ line correspond to present bounds as well as future sensitivities of EDM and SGWB observations, assuming $n=4$. As the right panel of Fig. \ref{fig:app1} shows for large hierarchy of scales $f_g/f_a=10^4$, a large part of the KSVZ-predicted $g_{a\gamma}-m_a$ parameter space including the point satisfying DM relic is already ruled out by neutron EDM and LIGO-O3 bounds. For smaller hierarchy $f_g/f_a=10^2$, the DM relic satisfying point remains currently allowed, but can be probed at future EDM experiments.


\providecommand{\href}[2]{#2}\begingroup\raggedright\endgroup

\end{document}